\documentclass[a4paper,11pt]{article}
\pdfoutput=1 

\usepackage{jinstpub} 

\usepackage{lineno}
\usepackage{adjustbox}
\usepackage{subfigure}
\usepackage{multirow}
\usepackage{color}

\title{\boldmath Lepton identification performance in Jets at a future electron positron Higgs Z factory}

\author[a]{Dan Yu,}
\author[b]{Taifan Zheng,}
\author[a, *]{Manqi Ruan}

\affiliation[a]{IHEP, China}
\affiliation[b]{Nanjing University, China}

\emailAdd{ruanmq@ihep.ac.cn}

\abstract{

Identifying the leptons inside jets is critical for the measurements of Higgs boson and the flavor physics program at the CEPC, a proposed Higgs/Z factory with the main ring circumference of 100 km. 
Using the CEPC baseline software, we analyze the identification performance of leptons generated inside a jet.
The jet leptons are identified with typical efficiency and mis-identification rate of 98\% and 1\% for energy higher than 2 GeV, with the $Z\to bb$ process at 91.2 GeV center-of-mass energy.
At the benchmark channel of the CEPC flavor program of $B_{c}\to\tau\nu$ with $\tau\to e\nu\nu$, the electrons are identified with inclusive efficiency times purity of 97\%, providing sufficient signal selection for the physics potential study. 
Compared to the single leptons, we found that these jet leptons identification efficiency degrades about 1-3\%, and the mis-identification rate increases by less than 1\% at the same working point.
We analyze this effect and found that it is mainly caused by defects in the particle flow reconstruction code in the calorimeters. The dependence between the performance of the shower reconstruction and the lepton identification is quantified.

}

\keywords{Particle identification methods, Performance of High Energy Physics Detectors, Simulation methods and programs}

\begin{document}
\maketitle
\flushbottom
\section{Introduction}

	The discovery of the Higgs boson at LHC completes the Standard model particle spectrum and provides a very sensitive probe to the fundamental physics principles underlying the Standard Model.	
	The electron-positron Higgs factories have low backgrounds and can make the absolute measurement of Higgs boson coupling; they are regarded as the top priority of the next collider, according to European Strategy and Physics Briefing Book \cite{EUStr,Brief}.
	Several electron-positron Higgs factories with better accuracy on the measurements of Higgs boson have been proposed, including the International Linear Collider (ILC)\cite{ilc}, the Compact LInear Collider (CLIC)\cite{clic}, the Future e+e- Circular Collider (FCC-ee)\cite{fccee} and the circular electron-positron collider (CEPC)\cite{cepc}. 
	Taking the CEPC as an example, it is expected to deliver 1 million Higgs bosons during its Higgs operation. The Higgs couplings will be measured to percent or even per mille level accuracy\cite{cepcAcccdr}.
		
	The CEPC will also operate at 91 GeV, producing nearly $10^{12}$ Z bosons\cite{cepc}. 
	Such a high luminosity Z factory provides unique opportunities for various flavor measurements.
	The CEPC Z factory can measure many of the rare decays of b/c hadrons and taus with precision beyond any ongoing or planned experiments.
	In addition, it would allow measurements of flavor violating Z decays with unprecedented precisions as a probe of new physics, such as the existence of sterile neutral fermions.
	The CEPC Z factory is complementary to existing B factories such as the LHCb and the Belle II.
	The environment of the CEPC Z factory is much cleaner, which provides smaller background than the LHCb. 
	The boost of b hadrons in the CEPC is much larger at the Z pole than the Belle II, which leads to larger displacement of secondary vertices and more collimated decay products.

\begin{figure}[htbp]
\centering
\includegraphics[width=.45 \textwidth,clip]{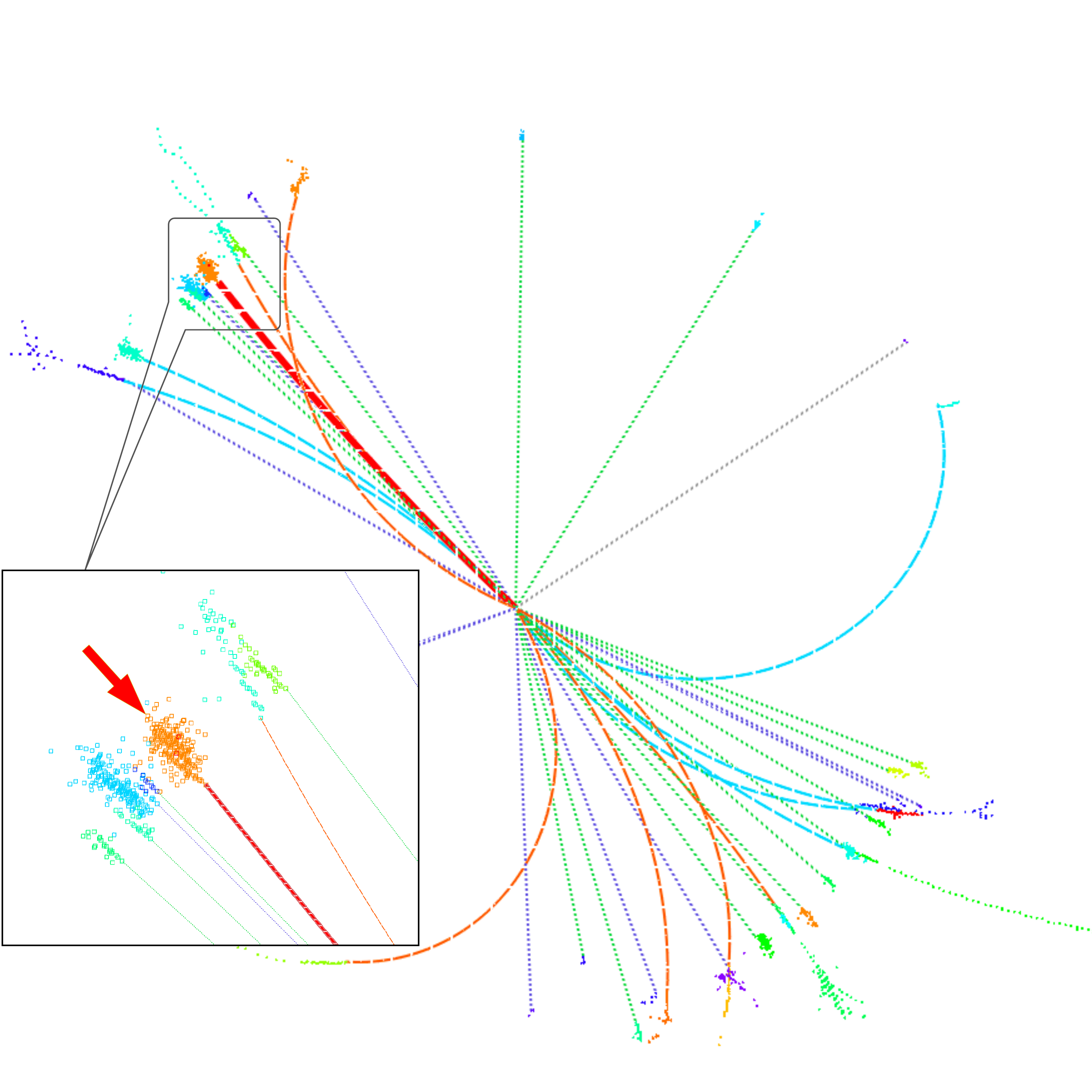} 
\label{ED_id}
\caption{An event display of an electron inside jet in $Z\to bb$ event, the arrow indicate the successfully reconstructed electron.}
\end{figure}

\begin{figure*}[htbp]
\centering
\includegraphics[width=.875\textwidth,clip,trim=0 0 0 0]{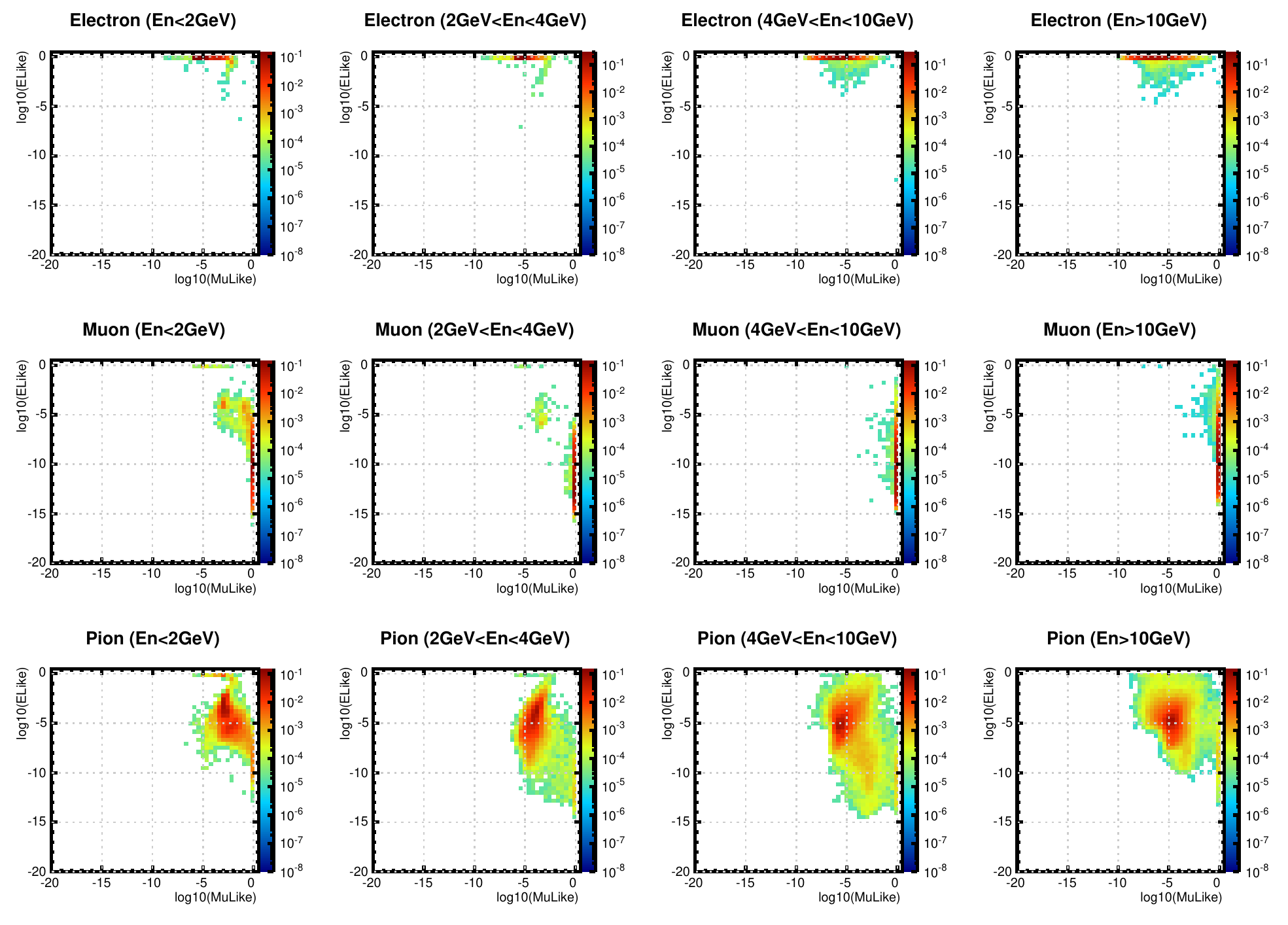} 
\caption{The E-likelihood and Mu-likelihood of single electrons, muons, and pions}
\label{SinglePhaseSpace}
\end{figure*}
 
\begin{figure*}[htbp]
\centering
\includegraphics[width=.875 \textwidth,clip,trim=0 0 0 0]{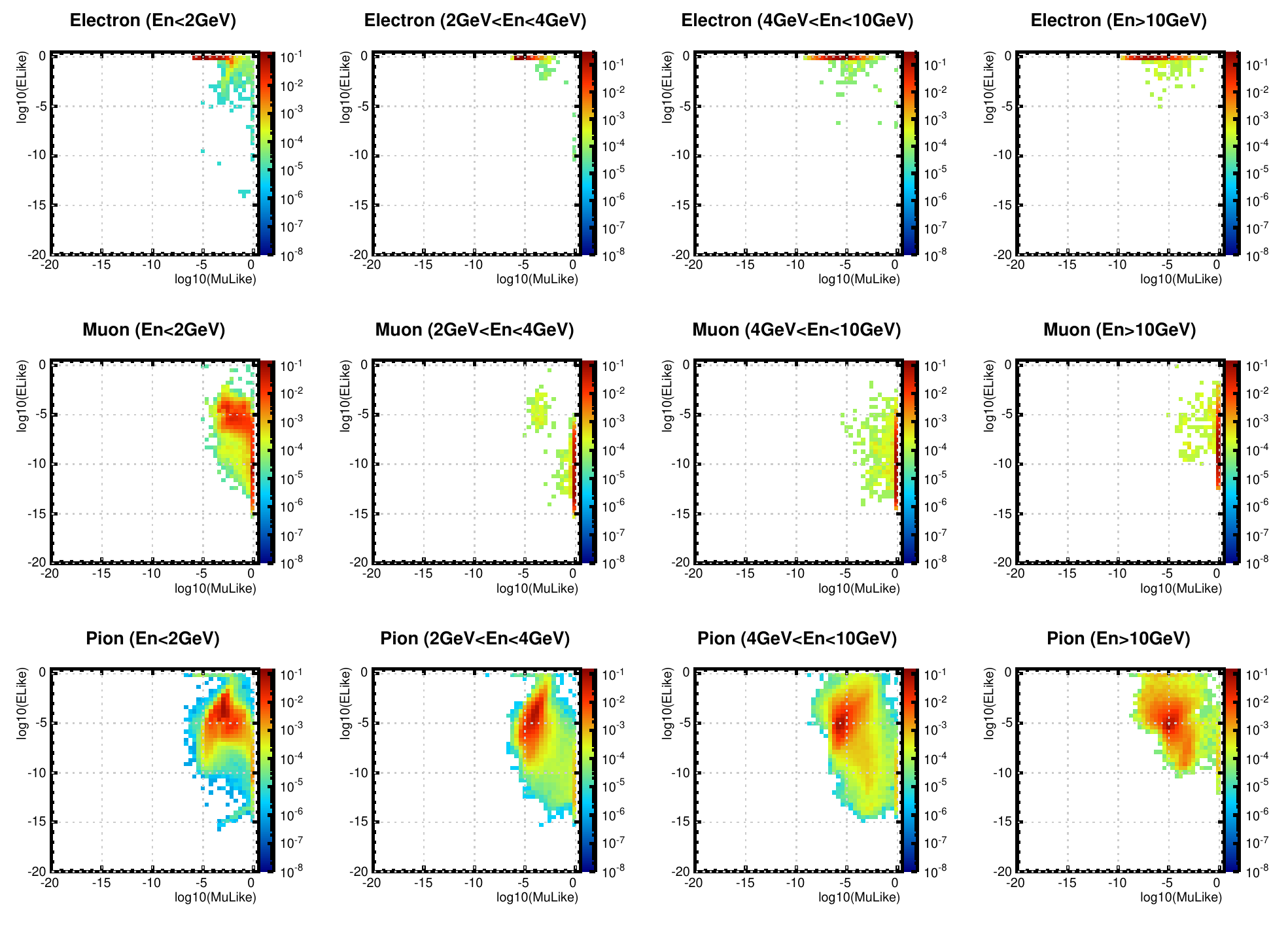} 
\caption{The E-likelihood and Mu-likelihood of electrons, muons, and pions in the jets}
\label{2d-likelihood}
\end{figure*}

	The lepton identification is essential for the physics measurements at the CEPC.  
	In high energy physics experiments, the leptons could be classified into leptons in jet (jet leptons for short in the following) and isolated ones.
	The jet leptons are usually QCD fragmentations of quarks and gluons or from the leptonic or semi-leptonic decay of heavy mesons in the events with jets.
	In these events, the lepton identification provides important input for the jet flavor tagging and the jet charge measurement in the analysis of Higgs coupling to $b/c/g$ and the $A_{FB}(b)$ in Z factory.
	The lepton identification is also important to tag the semi-leptonic decay of massive hadrons for the flavor physics analysis, such as the $R_{D} (\frac{BR(B\to D^(*)\tau\nu)}{BR(B\to D^(*)l\nu)})$ or $R_{K} (\frac{BR(B\to K^(*)\mu^{+}\mu^{-})}{BR(B\to K^(*)e^{+}e^{-}))})$ measurement.
	The isolated leptons are usually the prompt lepton, for instance, the initial leptons accompanied with the Higgs boson in $ZH$ events.
	At the electron-positron collider, ZH events with Z decaying into a pair of leptons are regarded as the golden channel for the $g(HZZ)$ and Higgs mass measurement.
	In these analyses, isolated lepton identification is essential for signal selection.

	A lepton identification package called LICH has been developed for the detectors using high granularity calorimeters\cite{LICH}.
	Using the baseline detector geometry for the Circular Electron-Positron Collider (CEPC) and the single charged particle samples with energy larger than 2 GeV, the LICH identifies electrons/muons with the efficiency higher than 99.5\% and controls the mis-identification rate of hadrons to muons/electrons to better than 1\%/0.5\%. 
	Applied on the Higgs recoil mass analysis with the $eeH$ and the $\mu\mu H$ channel, the signal events are identified with efficiencies of 95\%/97\% by requesting a pair of prompt leptons, and the backgrounds from fake leptons are reduced to a negligible level.\cite{LICH}

	Despite the physics importance of jet lepton identification, its performance has not been investigated. 
	In this paper, we applied LICH on the jet lepton in $Z\to bb$ events. 
	The overall efficiency for the electron is about 99\%. 
	However, for muon with energy under 15 GeV, the efficiency is degrading.
	The mis-identification rate from hadrons to leptons is about 1\%.
	Compared to isolated cases, the performance degrades about 1-3\% with the same working point.
	This is mainly because the distances between final state particles are closer, and reconstructed particles are more likely to be mixed with each other, as shown in Figure \ref{ED_id}.
	This paper aims to investigate the correlation between the clustering performance and the lepton identification.

	This paper is organized as follows. 
	The detector geometry and the samples are presented in section 2. 
	In section 3, an overview of jet lepton identification performance is presented. 
	In section 4, the calorimeter clustering performance is introduced, the correlations between jet lepton identification performance and calorimeter clustering performance are explored. 
	In section 5, the results are summarized, and further investigation is discussed.
	
\section{Detectors and Samples}

	The Particle Flow Algorithm (PFA) has become a trend of detector design for the high energy physics experiment, including the CEPC.
	The principle of the PFA is to reconstruct every final state particle in the most suited sub-detectors and reconstruct all the physics objects on top of the final state particles. 
	Charged particles' momenta are measured in the tracking detector with excellent precision, photons' energies are measured in the ECAL (with energy resolution typically of $\sigma(E)/E \sim 0.16/\sqrt{E}$), and neutral hadrons' energies are obtained from the HCAL (with energy resolution of $\sigma(E)/E \sim 0.5/\sqrt{E}$).
	The PFA-oriented detectors have high efficiency in reconstructing physics objects such as leptons, jets, and missing energy. 
	The PFA also significantly improves the jet energy resolution since the charged particles, which contribute the majority of jet energy, are usually measured with much better accuracies in the trackers than in the calorimeters.
	With the reconstructed final state particle information, the CEPC PFA, Arbor\cite{arbor}, can measure the boson mass at a resolution of 3.8\%, providing a clear separation of W, Z, and the Higgs bosons.
	
	To reconstruct every final state particle, the CEPC employs highly-granular calorimeters. 
	In addition to cluster separation, highly-granular calorimeters also provide detailed spatial, energy, and even time information on the shower developments.
	An accurate interpretation of this recorded information will enhance the physics performance of the full detector.
	The CEPC detector design takes the ILD reference\cite{cepc,ilc} and is optimized for the CEPC collision environment.
	The tracking system in the CEPC baseline employs a TPC(Time Projection Chamber), providing dE/dx information for the particle identification.

	The CEPC baseline detector is implemented in the CEPC software.
	The samples used in this paper are fully simulated $Z \to bb$ events using Whizard v1.95\cite{whizard} (Pythia6 \cite{pythia}) and MokkaC\cite{mokka,mokkaC}, reconstructed with Arbor.
	Because of limited computing resources, only 400k events are simulated, leading to about 37k electrons, 25k muons, and 500k charged hadrons.
	In this analysis, they are used as the charged particles in jets.

\section{Current Jet Lepton Identification Performance}

	The LICH is the default lepton identification package in the CEPC software. 
	It takes individual reconstructed charged particles as input, extracts 24 discriminant variables for the lepton identification, and calculates the corresponding likelihood to an electron or a muon.
	These variables contain not only the calorimeter information, such as the energy and hits distribution, but also the information from the tracking system, the $dE/dx$.
	The E-likelihood and Mu-likelihood for isolated leptons and pions are shown in Figure \ref{SinglePhaseSpace}, while the jet leptons are shown in Figure \ref{2d-likelihood}. 
	In most cases, as shown in the figures, the electrons peak at higher E-likelihood region and hardly overlap with others, the muons peak at higher Mu-likelihood region with little overlaps with pions at low energy, and the bulk part of pion distributions are well separated from the lepton peaks except the tails at high  Mu-likelihood region. 
	The LICH does not take other charged hadrons into account, therefore the performances in this article are only taking pions as charged hadrons.
	
	The phase space spanned by these two likelihoods is divided into different domains, corresponding to different catagories of particles. 
	The partition for particles of different types can be adjusted according to the requirement of different physics analyses.
	In this paper, we first take the E-likelihood higher than 1/3 to be electrons, the Mu-likelihood higher than 1/3  in the rest to be muons, and the remain particles to be pions.
	
	In Figure \ref{2d-likelihood}, the distribution for the jet leptons is expanding compared with Figure \ref{SinglePhaseSpace}, especially for energy lower than 4 GeV.
	When the muon energy is below 4 GeV (first two plots of the second row), a significant number of muons are mixed into the pion region.
			
\begin{figure}[htbp]
\centering
\subfigure{
\includegraphics[width=.45 \textwidth,clip]{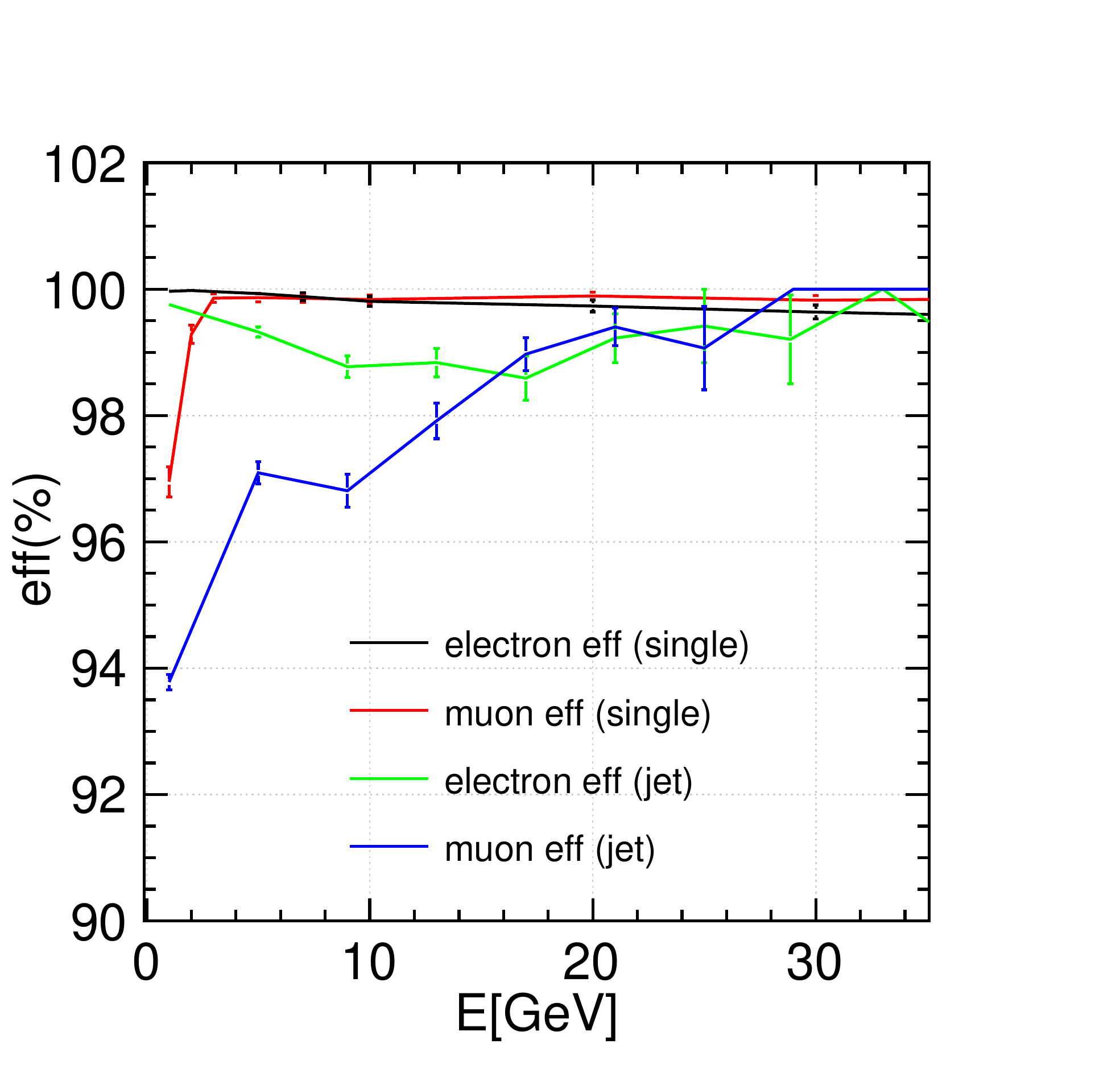} 
\label{Compare_all_eff}
}
\subfigure{
\includegraphics[width=.45 \textwidth,clip]{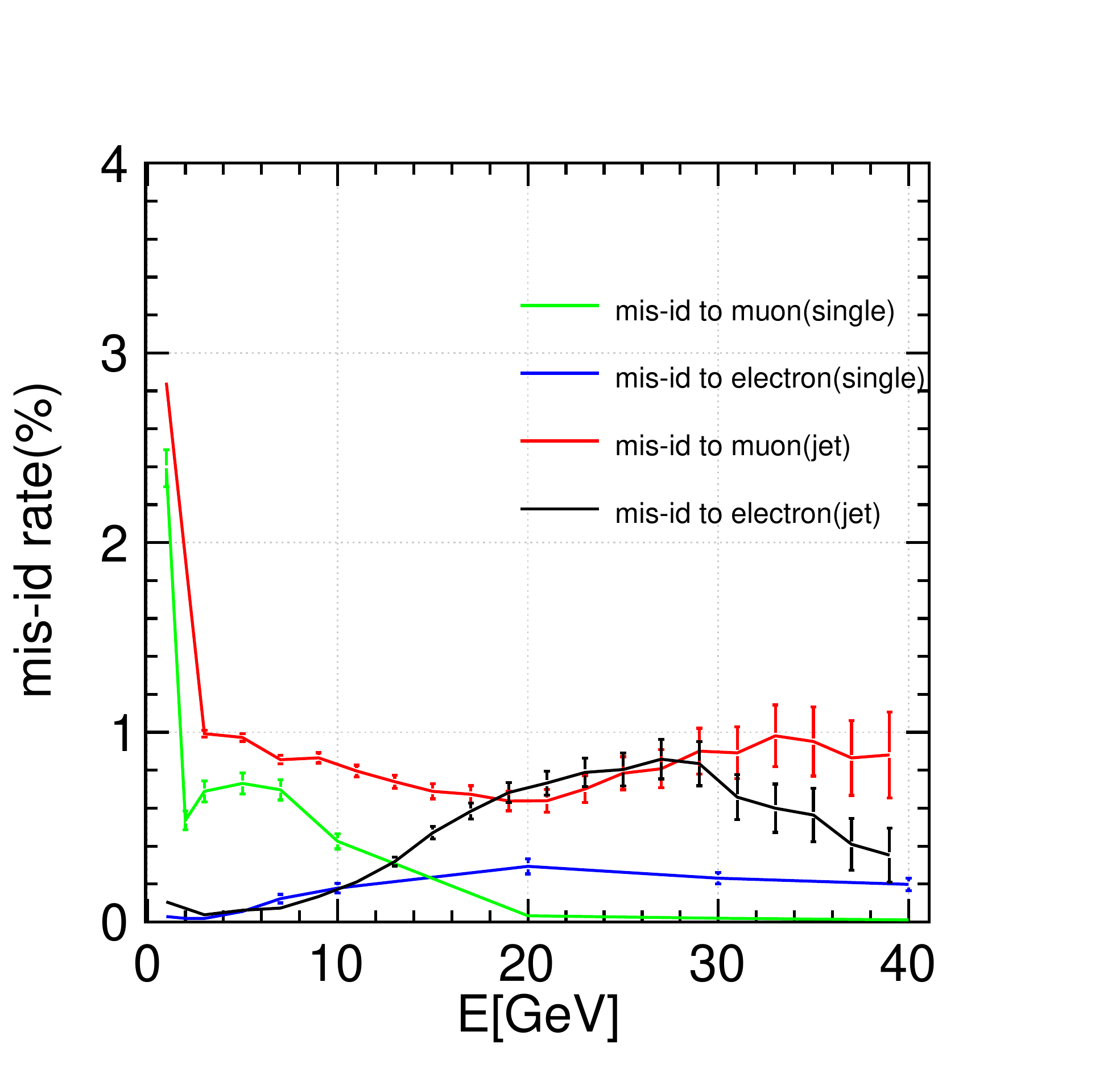} 
\label{Compare_all_mis}
}
\label{Compare_all}
\caption{The efficiency (up) and purity (down) of electrons and muons in jets, comparing with single particle case.}
\end{figure}

	Applying the same cut with single particle cases, we found that the efficiency and the mis-identification rate are degrading in jets, as shown in Figure \ref{Compare_all}. 
	Note that the minimum $P_t$ for the charged particle to reach the calorimeter barrel is about 1 GeV.
	 In order to show the performance of an entire energy range of (1-30) GeV, only the endcaps region is taken into account in this paper.
	The overall identification efficiencies for electrons and muons are 99\% and 98\%, respectively.
	For energy higher than 20 GeV, the efficiencies are comparable to the isolate performance.
	For lower energy, the muon efficiency degrades up to 3\%.
	The rate for charged hadrons to be mis-identified as electrons or muons increases by less than 1\% compared with the isolated case. 
	
\begin{figure}[htbp]
\centering
\includegraphics[width=.45 \textwidth]{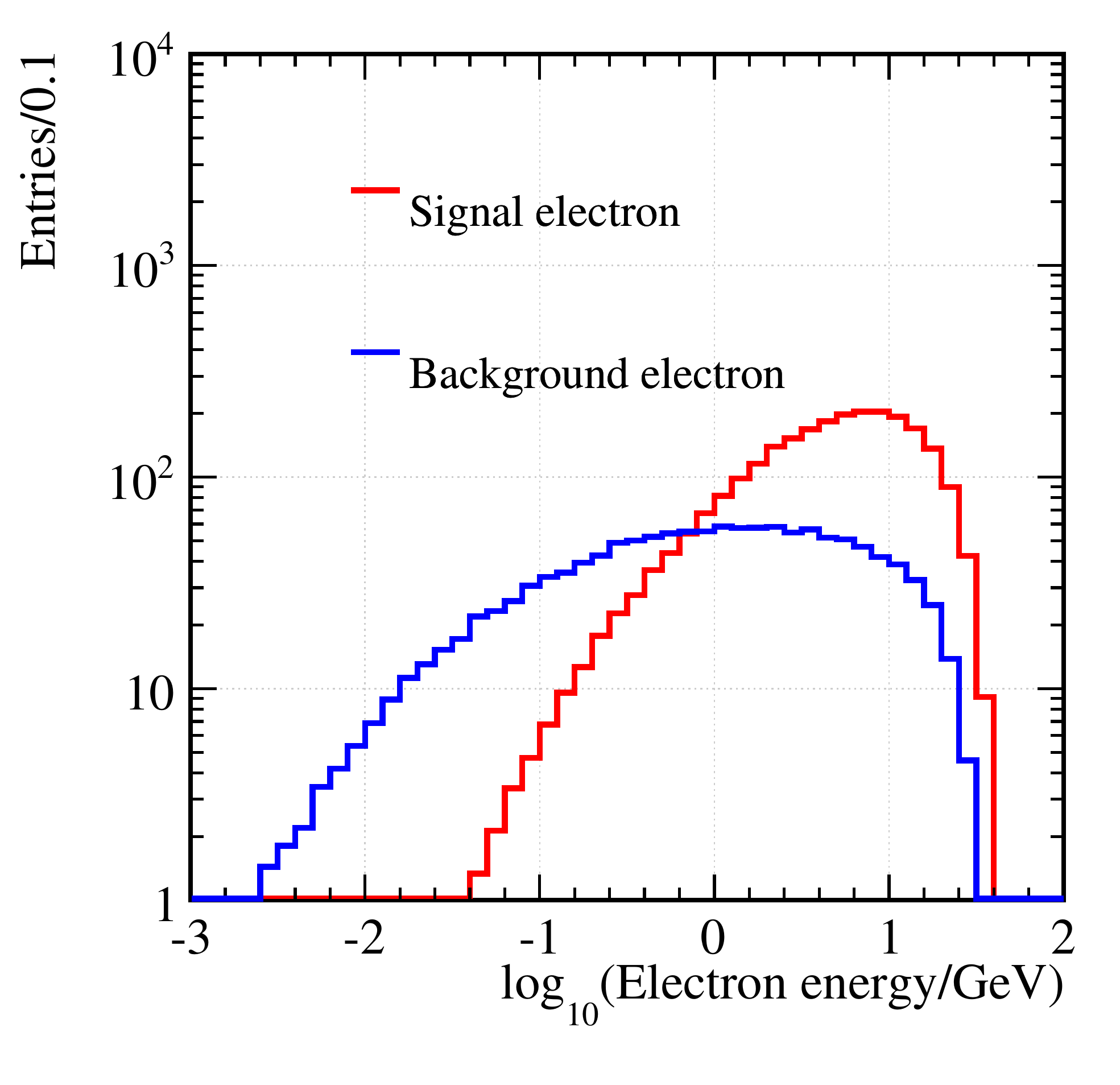} 
\caption{Energy of electrons in $B_{c}$ jets. Signal: electrons from $B_{c}\to \tau\nu$ with $\tau \to e\nu\nu$. Background: other electrons}
\label{EnSp_Bc}
\end{figure}

\begin{figure}[htbp]
\centering
\includegraphics[width=.45 \textwidth]{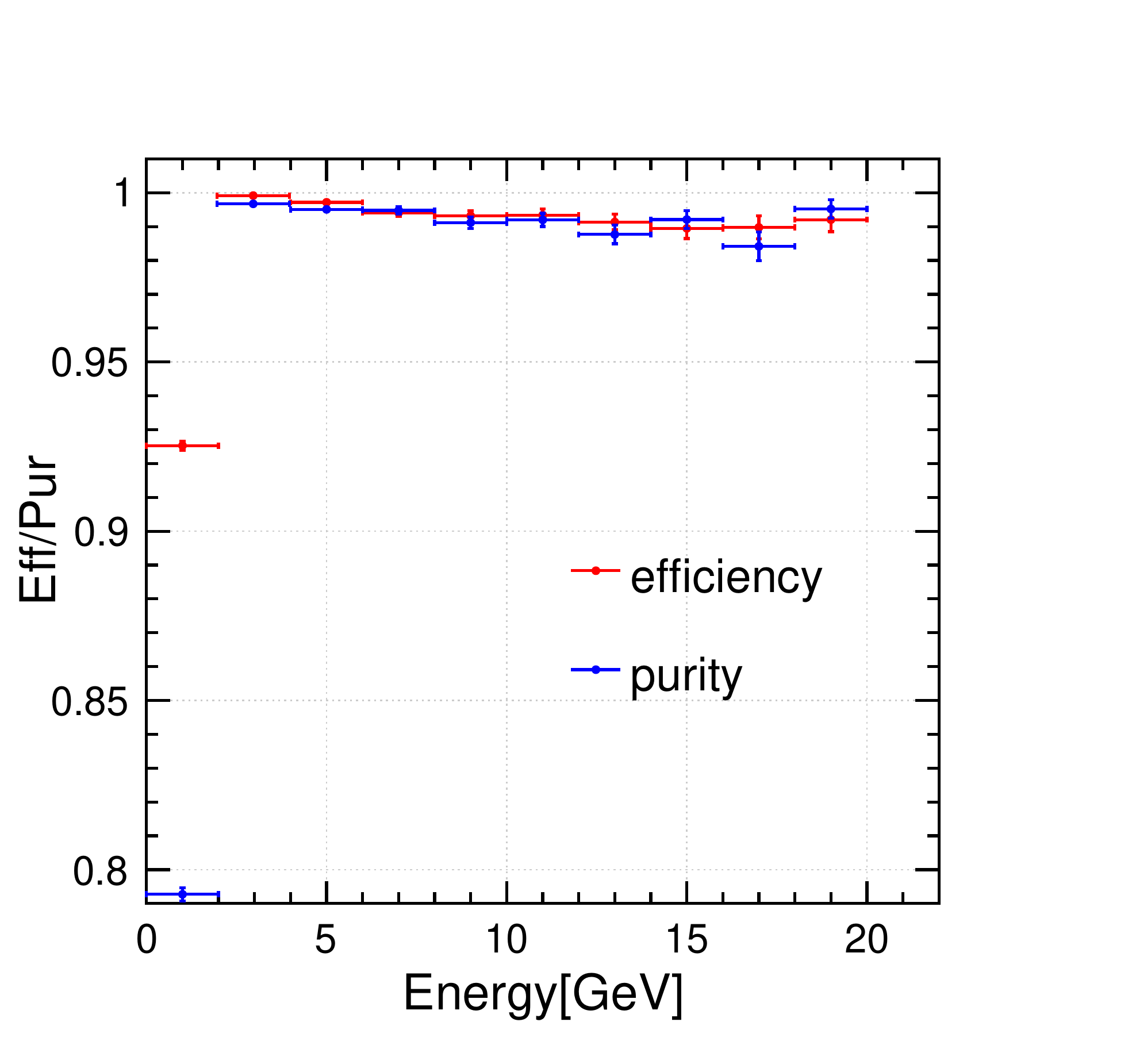} 
\caption{The identification efficiency and purity for electrons with different energy in $B_{c}$ jets}
\label{BcPerf}
\end{figure}

	The lepton identification is applied in the CEPC physics analysis. 
	Taking the $B_c \to \tau \nu$ analysis as an example, the energy spectrum for electrons in these events is shown in Figure \ref{EnSp_Bc} \cite{Bc_taifan}. 
Figure \ref{BcPerf} shows the electrons identification efficiency and purity in $B_c$ jets, and the overall efficiency times purity is 97\%.
Assuming the energy spectrum for muons in $\tau\to \mu\nu\nu$ channel is the same as the electrons in $\tau\to e\nu\nu$ channel, the muon identification efficiency times purity can be extrapolated to 72\%, using the particle statistics of $\tau\to e\nu\nu$ and the jet lepton identification performance in Figure \ref{Compare_all}.
However, for physics benchmarks with lower energies, we expect a larger degradation on the lepton identification performance.
	
\section{Diagnosis of performance}

As shown in the last section, the jet lepton identification efficiency is degrading 1-3\% compared with single particle identification.
Compared to the isolated leptons, the cluster of jet leptons are closer to other particles and are easier to be contaminated with each other and to lose efficiency, which might cause performance degradation.
In this section, we quantify the performance on calorimeter hit clustering and analyze the dependence between the clustering performance and the lepton identification.

\subsection{The efficiency and purity of the calorimeter cluster}

From the simulation, we can trace back the MC particle that contributes to a given calorimeter hit.
The clustering efficiency and purity are defined as follows, where $N_{correct}$ represents the number of hit number belonging to the correct MC particle, $N_{MC}$ is the total hit number of the MC particle in the detector, and $N_{clu}$ is the clustered hit number of the cluster.

\begin{equation}
\epsilon = N_{correct}/N_{MC},
p = N_{correct}/N_{clu}.
\end{equation}

\begin{figure*}[htbp]
\scriptsize
\smallskip
\centering
\includegraphics[width=.875 \textwidth]{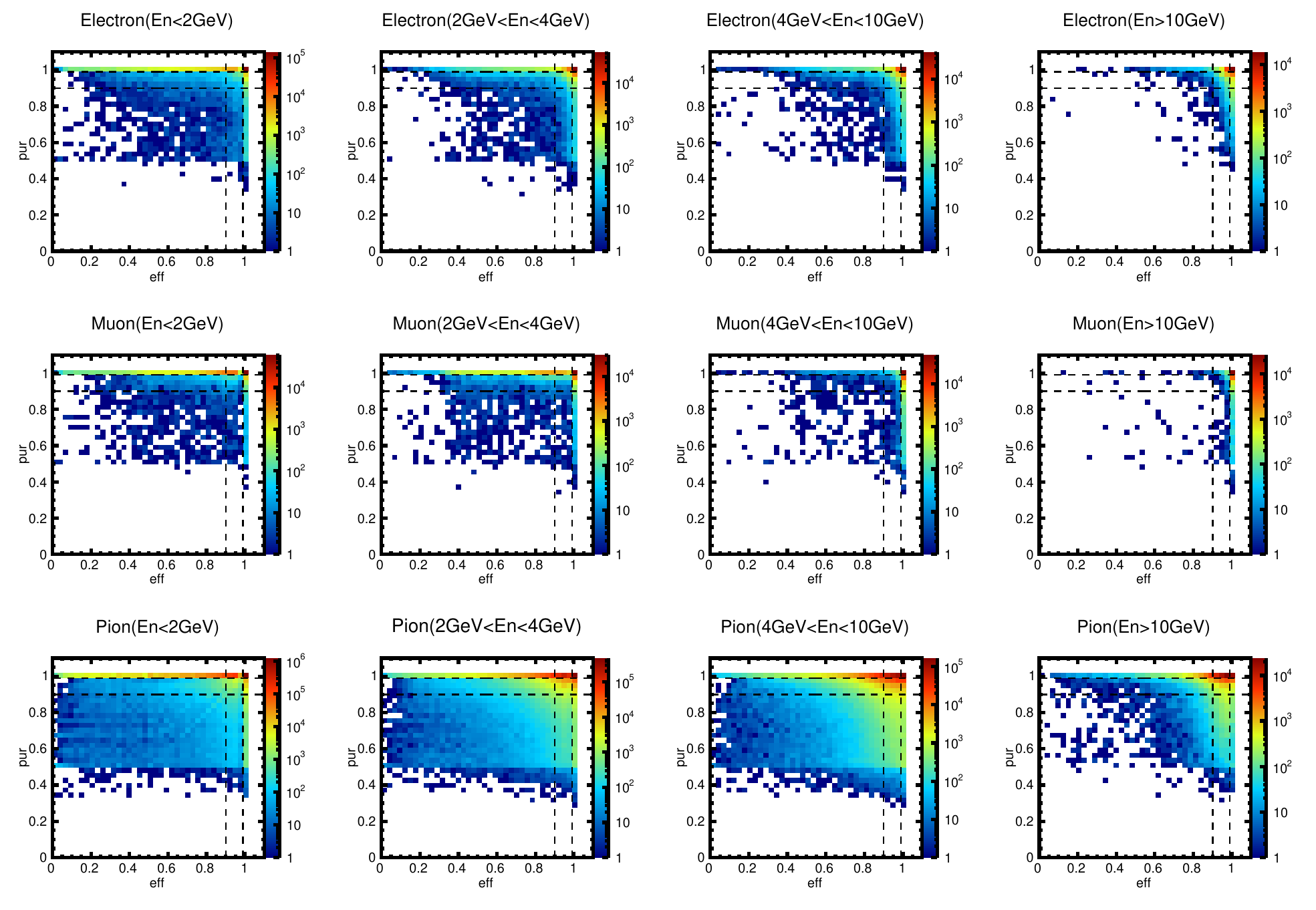} 
\caption{The clustering efficiency and purity of electrons and muons in jets}
\label{cluEffPurvsEn}
\end{figure*}

The clustering efficiency and purity of leptons and charged hadrons at different energies are shown in Figure \ref{cluEffPurvsEn}. 
It shows that the clustering performance distribution is more concentrated for higher energy.
This is because most lower energy electrons are from the $\pi_{0}$ Dalitz decay, while the higher energy ones are mostly decayed from the heavy mesons. 
The formers are more closed to each other than the latters.
For muons, the higher energy ones are more straight in the calorimeter, which is easier to be clustered.
The muons clustering performance is better than electrons and charged hadrons since its cluster is always a MIP(Minimum Ionization Particle) and not easy to be mixed with other clusters. 
From the last row of the Figure, the efficiency for charged hadron clusters are lower since these clusters are more likely to lose hits (splitting clusters). 
In general, 2/3 electromagnetic showers, 4/5 MIP showers, and 1/2 hadronic showers have perfect clustering performance in this $Z\to bb$ sample reconstructed with the current Arbor at the CEPC baseline.
In the next section, the identification will be performed for clustering efficiency times purity peaking at 1, and the ratio of clustering performance for different particles with different energies is shown in Table \ref{ratioTab}.

\begin{table}[htbp]
\centering
\caption{\label{ratioTab}The proportion of different clustering performance at different energies} 
\smallskip
\begin{tabular}{ccccc}
\hline
\multirow{2}{*}{particle} & \multirow{2}{*}{energy} & \multicolumn{3}{c}{ Clustering $\epsilon \cdot p$(\%)  } \\
\cline{3-5}
 &  & $= 1$ & $0.9-1$& $ <0.9$  \\
\hline
\multirow{4}{*}{$e$} & $<2 GeV$ & 71.06 & 8.45& 20.49 \\
&$2-4 GeV$ & 69.89 & 18.10 & 12.01 \\
&$4-10 GeV$ & 62.71 & 28.58 & 8.71\\
&$>10 GeV$ & 47.66 & 45.99 & 6.35 \\
\hline
\multirow{4}{*}{$\mu$} & $<2 GeV$ & 58.89 & 13.46 & 27.65 \\
&$2-4 GeV$ & 56.52 & 14.29 & 29.19 \\
&$4-10 GeV$ & 81.99 & 13.15 & 4.86 \\
&$>10 GeV$ & 83.19  & 13.67 & 3.14 \\
\hline
\multirow{4}{*}{$\pi$} & $<2 GeV$ & 57.53 & 15.99 & 26.48  \\
&$2-4 GeV$ & 36.67 & 34.77 & 28.56 \\
&$4-10 GeV$ & 19.94  & 45.47 & 34.59 \\
&$>10 GeV$ & 13.19 & 61.33& 25.47 \\
\hline
\end{tabular}
\end{table}

\subsection{Lepton identification at different clustering performance}

\begin{figure*}[htbp]
\centering
\includegraphics[width=.875 \textwidth,clip,trim=0 0 0 0]{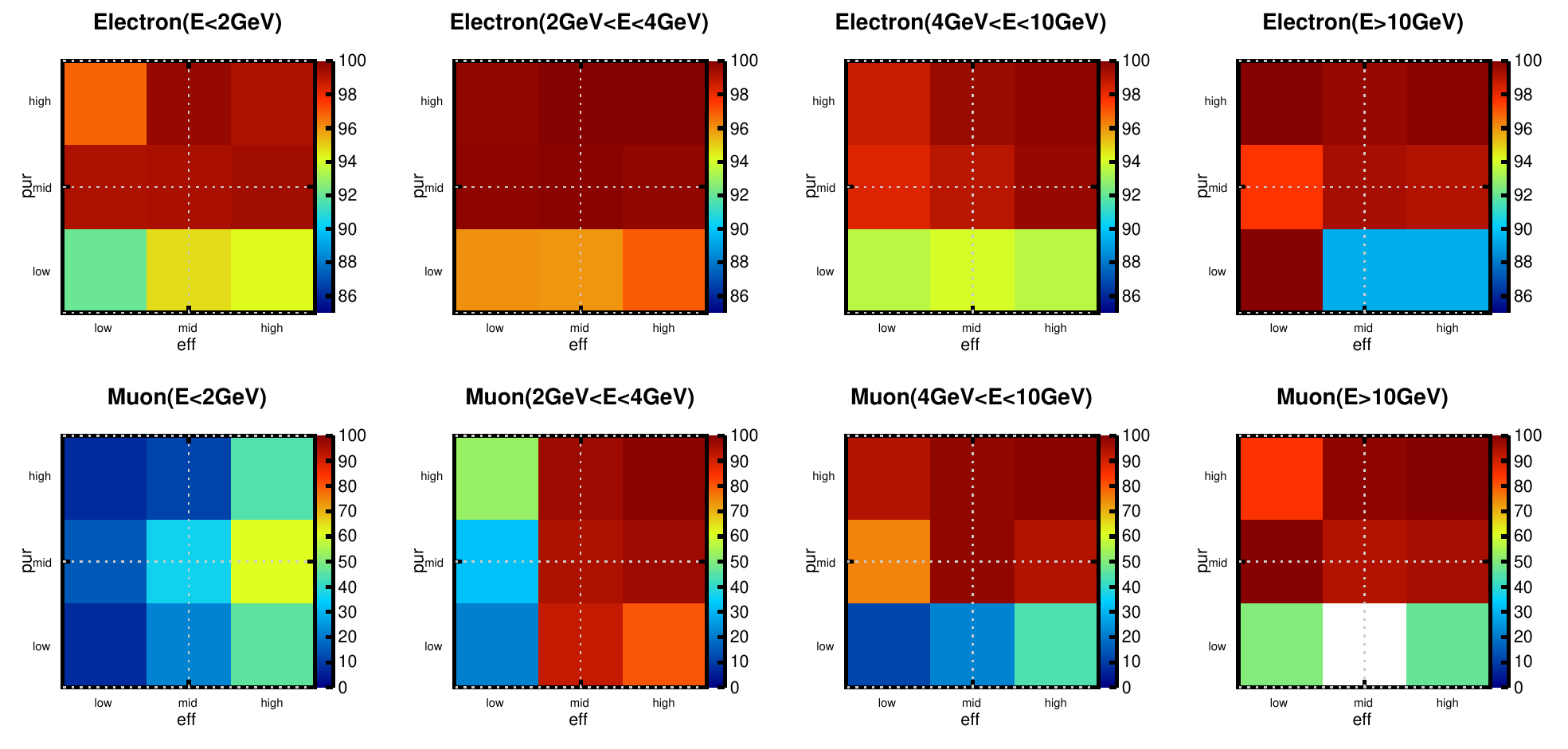} 
\caption{The lepton identification efficiency in jets depending on clustering efficiency and purity. The categories are defined as: high (=1), mid (0.9-1), low (<0.9).  The range for electron is from 85 to 100, for muon it is 0 to 100.}
\label{pidEffvsclu}
\end{figure*}

To investigate the correlation between lepton identification performance and the calorimeter clustering performance, the lepton identification efficiency is shown at different clustering performances in Figure \ref{pidEffvsclu}. 
Here each region corresponds to the spaces separated by the dash lines in Figure \ref{cluEffPurvsEn}.

For electrons, the PID (particle identification) efficiency is always higher than 90\% for all the clustering performance, even at low energies, where clustering is difficult, but the dE/dx plays the dominant role in identification.
The clustering efficiency affects the electron identification less than the clustering purity since the electron clusters are always compact and the splitting clusters are still electron-like.
On the other hand, the electron clusters are more likely to be mixed with EM showers, leading to another EM-like shower. This is relatively harmless to the PID efficiency but will reduce the purity.
However, if the clustering purity is too low (<0.4), the cluster will be less compact, leading to a higher probability of being identified as a hadron. 

For muons, the efficiency is lower while the clustering efficiency goes down.
The muons are always track-like in the calorimeters that go through the whole detectors. 
If the muon clusters are split, they are no longer an entire track passing through the calorimeter, which leads to lower muon likelihood.
The clustering purity becomes more critical as the energy surpasses 4 GeV since the muon cluster is no longer MIP-like if it is mixed with other hits.

\begin{figure*}[htbp]
\centering
\includegraphics[width=.875\textwidth,clip,trim=0 0 0 0]{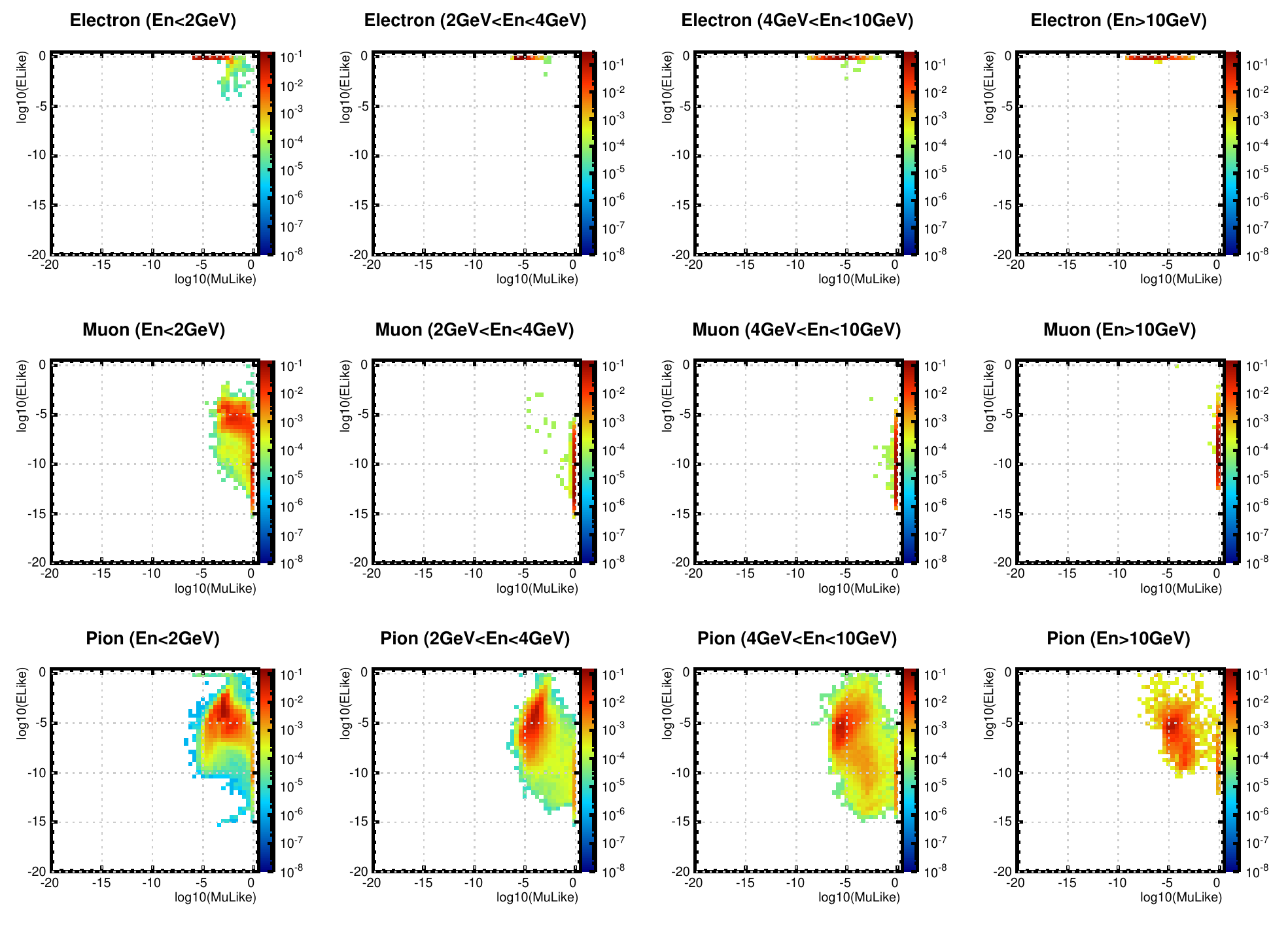} 
\caption{The Elikelihood and Mulikelihood of electrons, muons, and pions in the jets for clustering efficiency times purity equal to 1}
\label{Likeliness_high1}
\end{figure*}

\begin{figure*}[htbp]
\centering
\includegraphics[width=.875 \textwidth,clip,trim=0 0 0 0]{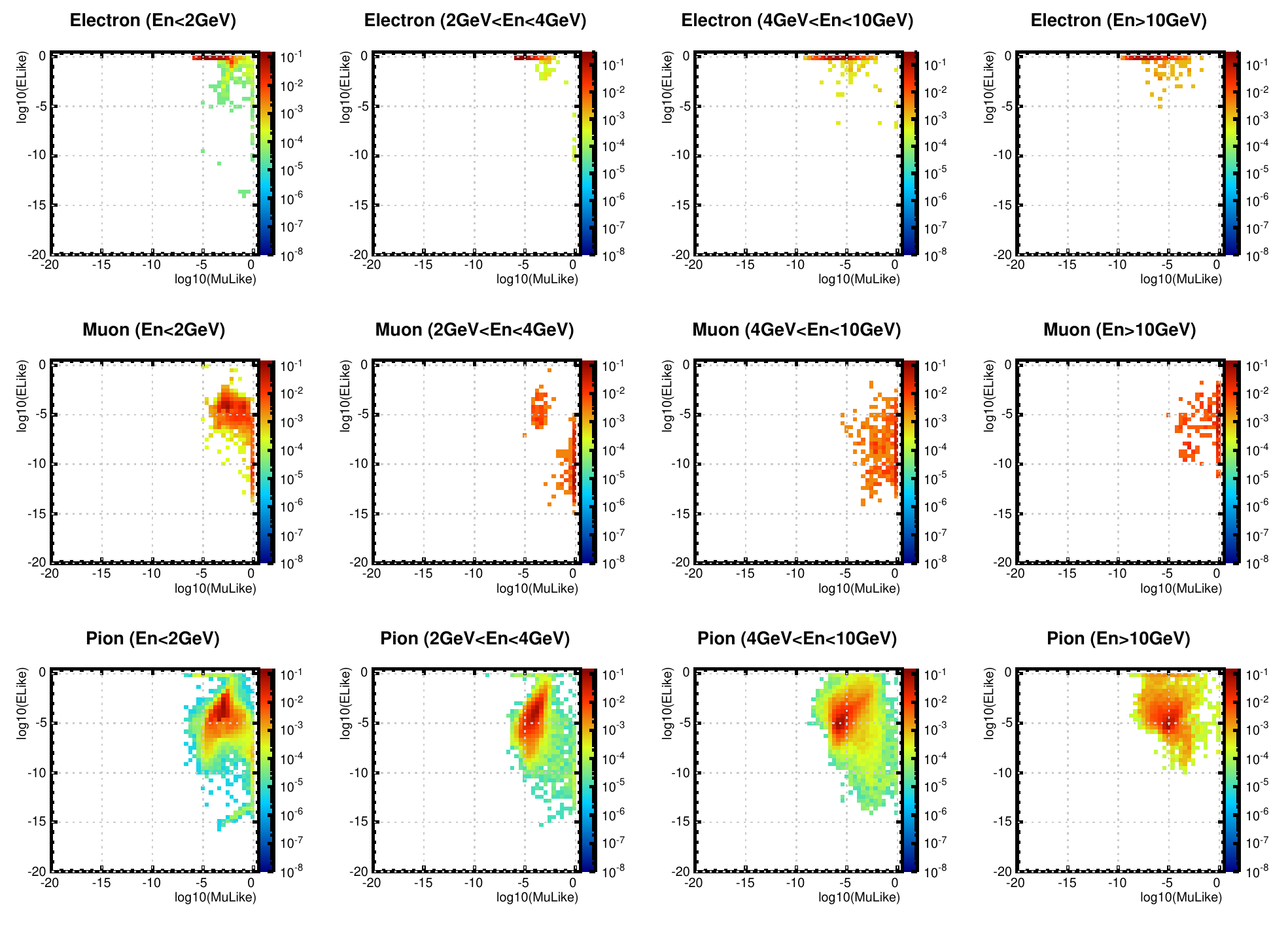} 
\caption{The Elikelihood and Mulikelihood of electrons, muons, and pions in the jets for clustering efficiency times purity less than 0.9}
\label{Lliness_low1}
\end{figure*}

Figure \ref{Likeliness_high1} and Figure \ref{Lliness_low1} show the likelihood distributions for clustering efficiency times purity equals one and less than 0.9.
The distribution expands when the clustering efficiency and purity are lower.
For higher energy leptons (last column), the distributions have much more significant tails for worse clustering, while the distributions for perfect clustering always peak with few fragments.
For muons with energy smaller than 4 GeV (the first two plots of the second row), the muon likeliness is degrading.
A considerable part of muons is at the pion region in the phase space.
However, for pions (the last row in Figure \ref{Likeliness_high1} and Figure \ref{Lliness_low1}), the distribution is drifting to the electrons when the clustering performance degrades.
A pion cluster is likely to be an EM cluster with some branches, and the EM part is more likely to be split with the others. 
This means that a pion is more likely to be mis-identified as an electron for lower clustering efficiency.
In this case, for better clustering performance, the rate of a pion to be mis-identified as a muon is higher instead of lower.
For perfect clustering (the last rows in Figure \ref{Likeliness_high1}), the pion distribution peaks at the muon region, which degrades the identification performance. 

\begin{figure}[htbp]
\centering
\includegraphics[width=.45 \textwidth]{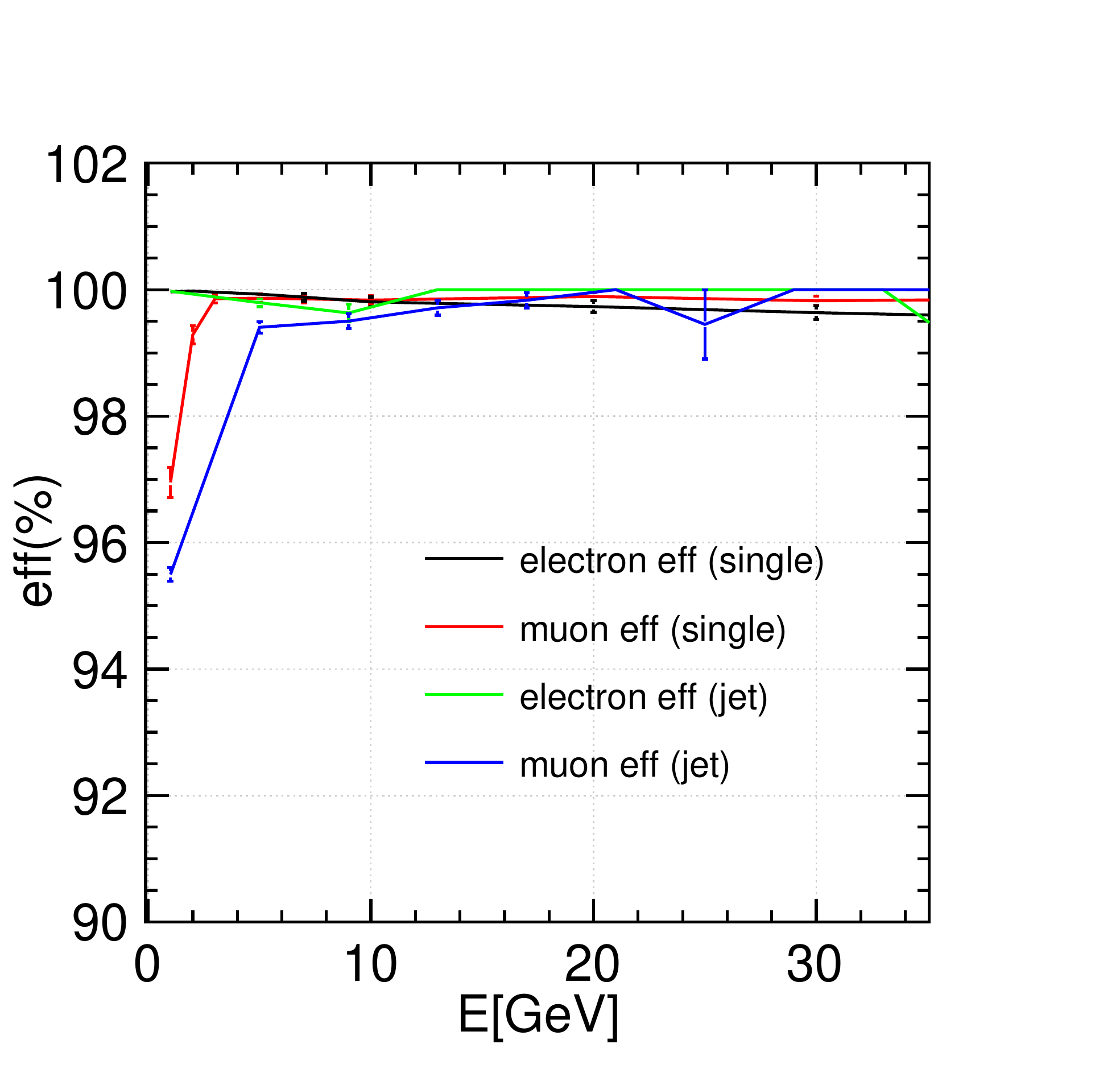} 
\caption{The lepton identification efficiency in jets at the best clustering efficiency and purity, compared with the single particle performance.}
\label{BestEff}
\end{figure}

With the clustering performance defined, the lepton identification in jets is modeled without the defects of clustering. 
While the cut on likeliness phase space keeps the same, the distributions for leptons with different clustering performance are changing, leading to different efficiencies and purities.
Figure \ref{BestEff} is the lepton identification performance when clustering efficiency and clustering purity equal 1, compared with the single particle events. 
At clustering efficiency times purity equals one, the lepton identification in jet converges to isolated lepton cases.

The clustering performance also describes the reconstruction results of different particles. 
For muons, once the energy of muon is larger than 3 GeV, the efficiency and purity peak at 100\%. 
For pions, the splitting is the main defect of clustering. 

	\section{Conclusion}

The identification of jet leptons is critical for the CEPC physics program: it is an essential input for the jet flavor and charge measurements, and it characterizes the physics processes with leptonic or semi-leptonic decays of the heavy flavor hadrons, etc.
Those jet leptons are surrounded by other final state particles. 
Thus it is essential to separate the detector hits induced by each final state particle.
This requirement is addressed by the Particle Flow oriented design at the CEPC baseline detector, which applies an advanced Particle Flow algorithm on a high-granularity calorimeter system.

Applying the baseline reconstruction algorithms at the fully simulated $Z\to bb$ process, we observe the overall efficiency of lepton identification in jets.
Thanks to the dE/dx provided by the TPC, the electron performance degrades at the percentage level.
However, muon identification is more challenging, especially at low energy.
Compared with isolated cases, the jet muon identification efficiency degrades up to 3\%.
The mis-identification rate of the pion to muon increases by about 0.5\% for energy smaller than 10 GeV.
At the  benchmark physics channel of $B_{c} \to \tau\nu$ with $\tau\to e\nu\nu$, the overall efficiency and purity are shown in Table \ref{Bc_tab}.
The estimation for $\tau\to \mu\nu\nu$ channel performance is also listed.
The low purity is mainly limited by the rate of pions mis-identified as muons and the high statistics of low energy pions.
Since the jet lepton performance is comparable to isolated lepton cases when the clusters are correctly reconstructed, we extrapolate the lepton identification efficiency times purity in the $B_{c} \to \tau\nu$ channel using the isolated lepton performance, as shown in the table.

\begin{table}[htbp]
\centering
\caption{\label{Bc_tab}The lepton identification performance for $B_{c} \to \tau\nu$ with $\tau\to e\nu\nu$ and $\tau\to \mu\nu\nu$ with current reconstruction and estimated perfect clustering using isolated lepton performance} 
\smallskip
\begin{tabular}{ccccc}
\hline
& &  $e$ in $\tau\to e\nu\nu$ & $\mu$ in $\tau\to \mu\nu\nu$ \\
\hline
\multirow{3}{*}{current reconstruction} & $\epsilon$ & 99.5\% & 95.3\% \\
&$p$ & 98.0\% & 75.5\% \\
&$\epsilon \cdot p$ & 97.5\% & 71.9\% \\
\hline
\multirow{3}{*}{perfect clustering} & $\epsilon$ & 99.9\% & 99.3\% \\
&$p$ & 99.4\% & 86.4\% \\
&$\epsilon \cdot p$ & 99.3\% & 85.8\% \\
\hline
\end{tabular}
\end{table}

From the comparison, the jet lepton identification can be significantly improved with an optimized cluster reconstruction.
For other physics analyses with leptons at even lower energy or event with multiple muons, this improvement is more required.
In this paper, the reconstruction performance is quantified using clustering efficiency and purity, and the dependence of the jet lepton identification on this performance is investigated.
This characteristic of clustering efficiency and purity can also be used as an adequate tool to evaluate the performance of calorimeter and associated reconstruction. 

One thing worth noting is that the muon system is not yet included in LICH. An improvement is expected if the muon detector information is applied in future studies.

\acknowledgments

The authors would like to thank Chengdong FU for providing the simulation tools. 
This study was supported by the National Key Program for S\&T Research and Development (Grant NO.: 2016YFA0400400),
the Hundred Talent programs of Chinese Academy of Science No. Y3515540U1. 

The work is supported by the Beijing Municipal Science \& Technology Commission, project No. Z191100007219010.


\end{document}